\documentclass[twocolumn,showpacs,preprintnumbers,amsmath,amssymb,superscriptaddress]{revtex4}
\divide\topmargin by 3
\usepackage{graphicx}
\usepackage{dcolumn}
\usepackage{bm}

\begin{document}

\title{\textit Impurity-induced tuning of quantum well states in spin-dependent
resonant tunneling}
\author{Alan Kalitsov}
\affiliation{Department of Physics, California State University,
Northridge, CA 91330-8268}
\author{A. Coho}
\affiliation{Department of Physics, California State University,
Northridge, CA 91330-8268}
\author{Nicholas Kioussis}
\email[E-mail me at: ]{nick.kioussis@csun.edu.}
 \affiliation{Department of Physics,
California State University, Northridge, CA 91330-8268}
\author{Anatoly Vedyayev}
\affiliation{Faculty of Physics, Lomonosov Moscow State University, 119992 Moscow, Russia}
\author{M. Chshiev}
\affiliation{Department of Physics, Virginia Polytechnic Institute and State University, Blacksburg, VA 24061-0435 }
\author{A. Granovsky}
\affiliation{Faculty of Physics, Lomonosov Moscow State
University, 119992 Moscow, Russia}

\date{\today}
\begin{abstract}
{We report exact model calculations of the spin-dependent
tunneling in  double magnetic tunnel junctions in the presence of
impurities in the well. We show that the impurity can tune
selectively the spin channels giving rise to a wide variety of
interesting and novel transport phenomena. The tunneling
magnetoresistance, the spin polarization and the local current can
be dramatically enhanced or suppressed by impurities. The
underlying mechanism is the impurity-induced shift of the quantum
well states (QWS) which depends on the impurity potential, impurity
position and the symmetry of the QWS.}
\end{abstract}
\maketitle

Tunneling of spin-polarized electrons through
magnetic tunnel junctions (MTJ) has attracted\cite{thornton} wide and sustained
interest in the past few years, both experimentally
\cite{moodera2, tsymbal2,moodera1,yuasa,moodera3} and
theoretically\cite{tsymbal2,petukhov,butler,chshiev}.
This is due to the potential applications of MTJ in
spin-electronic devices, such as magnetic sensors and magnetic
random-access memories. The key
point for these applications is the tunnel magnetoresistance
(TMR), i.e. the dependence of the tunneling current on the
relative orientation of the magnetization of the ferromagnetic
layers, which can be changed by an applied magnetic field\cite{moodera3}.

Double magnetic tunnel junctions (DMTJ) consist of a central
metallic layer (quantum well) between two insulating barriers and
two ferromagnetic electrodes. The insulating layers are thin
enough for electrons to tunnel through the barriers if a bias
voltage is applied between the electrodes. The TMR behavior in
DMTJ is determined by quantum well states (QWS) formed in the
middle layer when a resonance condition is
fulfilled~\cite{yuasa}. The TMR can be dramatically enhanced when
spin-polarized electrons {\it resonantly} tunnel through the
middle layer~\cite{petukhov}. Theoretical formulations of the TMR in DMTJ are usually based on models which assume perfect systems
\cite{petukhov,chshiev,zheng,hayashi,wang,zhang,miyamoto}.
The TMR exhibits an amplitude-varying oscillatory behavior as a function of the
thickness of the middle layer with a period of $\pi/k^\sigma_F$,
where $k^\sigma_F$ is the spin-dependent Fermi wave vector in the
middle layer~\cite{zheng}. Chshiev {\it et al} introduced\cite{chshiev}
phenomenologically the electron scattering in the middle layer
without taking into account vertex corrections. Consequently, they
introduced an effective electric field within each barrier in order to
satisfy the continuity equation for the current~\cite{fisher}.

However, actual MTJ contain large amounts of disorder in the electrodes,
in the barriers, in the quantum well, and at the electrode/barrier
or electrode/quantum well interfaces\cite{tsymbal2, jansen}.
This disorder may represent interdiffusion at the interfaces, interface roughness,
 and impurities. Experiments in single MTJ have suggested that disorder can
 affect the TMR in a critical way giving rise to impurity-assisted tunneling \cite{tsymbal2,jansen}.
While the effect of impurities within a single barrier has been studied
theoretically recently~\cite{tsymbal1,vedyayev}, its role in DMTJ remains an
unexplored area thus far.


In this paper we present exact model calculations
of the spin-dependent resonant tunneling in double MTJ structures
to study for the first time the effect
 of magnetic and nonmagnetic impurities in the magnetic middle layer on the
TMR, spin polarization, and local current as a function of
external bias. This approach conserves the continuity of the
current in contrast to Ref. \cite{chshiev}. We show that the
impurities may induce a shift of the original QWS depending on the sign of the impurity scattering potential, the impurity position, and the symmetry of the original QWS. These
effects can tune selectively the spin channels giving rise to a
wide variety of novel and interesting spin-dependent transport phenomena,
such as a dramatic enhancement or suppression of the TMR and the
spin polarization, and a sign reversal of the spin polarization.
Interestingly, the calculations reveal that even though the effect of the impurity on the average spin current
is small for antisymmetric QWS, the local spin current exhibits
strong variation.


We employ the free-electron band model to describe the electron
tunneling in a double MTJ  with impurities in the metallic
middle layer of width $b$, whose scattering is modeled by a
spin-dependent $\delta$-function  scattering potential. Figure 1
shows a schematic diagram of the energy bands for the parallel
configuration under a small external bias for the majority and
minority spin carriers, respectively. The geometric parameters of the DMTJ are
chosen in such a way that the spin-dependent energies of the QWS,
$E^{\sigma}_{R}$, fall within the majority
(minority) spin band and they are of antisymmetric (symmetric)
character, respectively. The impurity potential $V^{\sigma}$ relative to the
bottom of the band in the middle layer is also shown with green
(red) line for positive (negative) sign, with the impurity placed
at the center of the well.

\begin{figure}
\includegraphics[width=3in,height=3in]{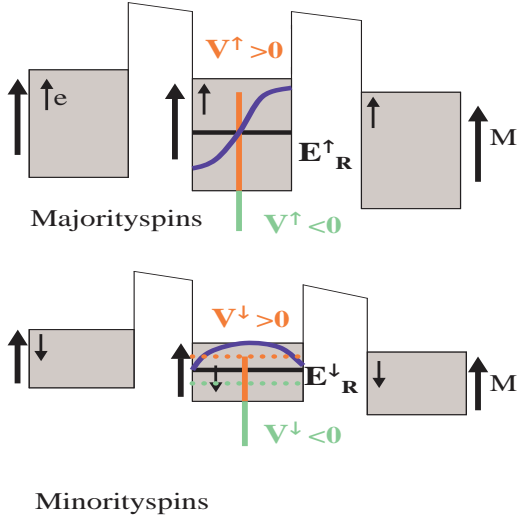}
\caption{\label{fig:epsart}  Schematic energy diagram for the parallel
configuration for the majority and minority spin carriers. The energies of the QWS
denoted by $E^{\sigma}_{R}$,  fall within the majority (minority)
band and they are antisymmetric (symmetric), respectively. The impurity $\delta$-type potential
$V^{\sigma}$ of positive (negative) sign relative to the bottom of
the band in the well is denoted with red (green) line,
respectively. There is no shift of the antisymmetric QWS, whereas
the symmetric QWS shift towards higher (lower) values if
$V^{\sigma} >0 (<0)$.}
\end{figure}

Employing the WKB approximation in the barrier region, the
one-electron Green's functions for the clean DMTJ are
solutions of the Schr$\ddot{o}$dinger equation,

\begin{eqnarray}
&& \left( E+\frac{\hbar ^2}{2m_i}\left( \frac{\partial^2}{\partial
z^2} -\kappa^2 \right) -\tilde{V}_i^\sigma -ev(z)
\right)G^\sigma_\kappa(z,z^\prime)= \nonumber \\ && =
\delta(z-z^{\prime}),
\end{eqnarray}

\noindent where $\tilde{V}_i^\sigma = V_i^\sigma$ in the metallic
layers, and $\tilde{V}_i^\sigma = U_i$ in the barriers. Here,
$V_i^\sigma$ is the spin-dependent potential (or the bottom of the
band) of the $i$-th metal, $U_i$ is the potential barrier height in the i{\it th} barrier, and $v_i(z)$ is the voltage drop within the i{\it
th} layer, assumed to be nonzero only in the barriers. The $z$ is
the coordinate perpendicular to the interface, $\kappa$ is the
in-plane wave vector of energy $E$, and $m_i$ is the electron
effective mass in the $i$-th layer. The coefficients of
the wave functions in each layer
are determined by the boundary conditions at the interfaces.

The one-electron Green's function in the presence of the impurity is determined from the Dyson equation:

\begin{eqnarray}
\tilde{G^\sigma}(\rho,z,\rho^{\prime},z^{\prime}) &&=G^\sigma(\rho,z,\rho^{\prime},z^{\prime})+G^\sigma(\rho,z,\rho_0,z_0)T^\sigma \times \nonumber \\
&& \times G^\sigma(\rho_0,z_0,\rho^{\prime},z^{\prime}),
\end{eqnarray}
\noindent where $T^{\sigma} = V^\sigma[1-V^\sigma
G^{\sigma}_{33}(\rho_0,z_0,\rho_0,z_0)]^{-1}$ is the T-matrix,
$V^\sigma=V_{imp}^\sigma~\delta (z-z_0) - V^\sigma_3$,  and
$\rho_0,z_0$ is the impurity position. Here, $V^\sigma$
($V_{imp}^\sigma$) refers to the
impurity potential relative to the
bottom of the band (Fermi energy) of the well. Depending on the
width of the quantum well and the sign of $V^\sigma$, the poles
of the T-matrix give rise to {\it impurity-induced} shift of resonances,
which can in turn enhance or suppress selectively the current
density of the majority or minority spin channel for a given
magnetization orientation.

The local current density for spin $\sigma$ is given by~\cite{duke}

\begin{eqnarray}
j^{\sigma}(\rho - \rho_0,z) &&=\frac{e}{\pi\hbar}\int \left[
f(E)-f(E+eV_{ext}) \right] \times \nonumber \\
&& \times D^{\sigma}(E,\rho - \rho_0,z)\,dE,
\end{eqnarray}

\noindent where $f(E)$ is the Fermi-Dirac distribution function and
the transmission probability
$D^\sigma$ is ~\cite{levy}
\begin{equation}
D^\sigma(E,\rho - \rho_0,z )=\left( \frac{\hbar ^2}{2m} \right) ^2
\sum_{k,k^\prime} A_{k,k\prime}^\sigma
\stackrel{\leftrightarrow}{\nabla}_z
\stackrel{\leftrightarrow}{\nabla}_{z^{\prime}}
A_{k,k\prime}^\sigma.
\end{equation}
Here, $\stackrel{\leftrightarrow}
{\nabla}_z=(1/2)(\stackrel{\rightarrow}{\nabla}_z-\stackrel{\leftarrow}{\nabla}_z)$
is the antisymmetric gradient operator,
$A_{k,k\prime}^\sigma=(1/2)[G_{k,k\prime}^{ret,\sigma}-G_{k,k\prime}^{adv,\sigma}]$,
where $G_{k,k\prime}^{ret,\sigma}$ and
$G_{k,k\prime}^{adv,\sigma}$ are the retarded and advanced Green's
functions, respectively. The total transmission probability $D = D^{(0)} +
D^{(1)} + D^{(2)}$, where $D^{(0)}$ is the transmission
probability in the absence of impurity, and $D^{(1)}$ and
$D^{(2)}$ are the transmission probabilities proportional to the
T-matrix and to the T-matrix squared, respectively. The
average current $j^\sigma$ can be calculated from Eq. (3) but with
$D^{\sigma}(E,\rho - \rho_0,z)$ replaced with its average value
\begin{equation}
<D^{\sigma}(E,z)> = \frac{N_{imp}}{N} \int D^\sigma(E, \rho -
\rho_0,z )d\rho ,
\end{equation}
where the number of impurities $N_{imp} = cN$, $c$ is  the uniform
impurity concentration, and $N$ is the total number of atoms in
the plane. It is important to note that the average current is
independent of $z$ satisfying the current continuity equation. In the following, we model an $Fe/Oxide/Fe/Oxide/Fe$ DMTJ, with
$V_1^\uparrow=V_3^\uparrow=V_5^\uparrow= -4.58eV$ and $V_1^\downarrow=V_3^\downarrow=V_5^\downarrow=-0.66eV$~\cite{moodera1}.
The barrier widths are 8~\AA and 2~\AA, respectively, the impurity
concentration  is $c=5\%$, and the impurity is placed at the
center of the well.

In Fig. 2 we show the $TMR = (j_P-j_{AP})/j_{AP}$\cite{moodera3}
as a function of the well thickness for the perfect DMTJ and
in the presence of non-magnetic impurities with
$V_{imp} = 0, +2, -2 ~eV$. Here, $j_P$ ($j_{AP}$)
is the current density in the parallel (antiparallel)
configuration in which the leads are ferromagnetically
(antiferromagnetically) aligned to the middle magnetic layer.
The peak in the TMR at b=4.3~\AA ~ for
the perfect DMTJ can be dramatically enhanced (suppressed)
by impurities with positive
(negative) $V^\sigma$ which can shift the original QWS. For
b=4.3~\AA, the original QWS for the majority (minority) spin for
the parallel (antiparallel) configuration are {\it below} $E_F$ and they
are antisymmetric. On the other hand, the QWS for the minority
(majority) spin for the parallel (antiparallel) configuration are
{\it symmetric} and {\it above} $E_F$. This is due to the
larger (smaller) $k_F^\sigma$ in the well for the majority
(minority) spin band~\cite{zheng}. The effect of the impurity on
the QWS can be understood in terms of a simple quantum mechanical
model, namely a $\delta$-function impurity-potential, $V^\sigma$,
within a potential well. Symmetric QWS are shifted towards
higher (lower) energies for positive (negative) impurity
potential. In contrast, antisymmetric QWS are not shifted.
Thus, the positive (negative) impurity-potential shifts the QWS
away from (closer to) the Fermi energy, suppressing (enhancing)
both $j_P^{\downarrow}$ and $j_{AP}^{\uparrow}$, and hence
increasing (decreasing) the TMR. On the other hand, the
$j_P^{\uparrow}$ and $j_{AP}^{\downarrow}$ are not affected by the
impurity.

\begin{figure}
\includegraphics[width=3in,height=3in]{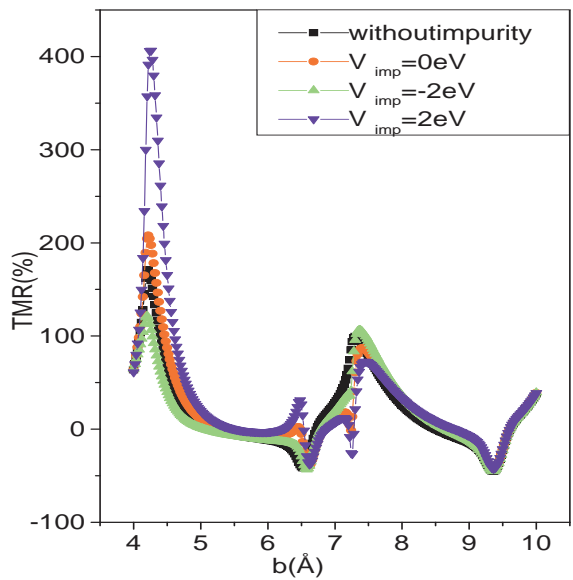}
\caption{\label{fig:epsart} TMR versus the middle layer thickness
$b$ in the presence or absence of nonmagnetic impurities for
different values of the impurity potential, $V_{imp}$.}
\end{figure}

For b= 7.24~\AA, in the absence of impurity the majority and
minority spin QWS for both the parallel and antiparallel
configurations are {\it symmetric} and {\it below} $E_F$. For $V_{imp}$ = +2,
0 eV, both the $V^\uparrow$ and $V^\downarrow$ are positive and the QWS are
shifted closer to E$_F$. Thus, the current for both parallel
and antiparallel configurations increases and the TMR
decreases. For $V_{imp}$ = - 2 eV the QWS for the minority
(majority) spin for the parallel (antiparallel) configuration are
lowered in energy and hence the corresponding currents decrease.
On the other hand, the QWS for the majority (minority) spin for
the parallel (antiparallel) configuration are raised in energy and
hence the corresponding currents increase. However, it turns out
these current components compensate each other and the TMR is not
altered much.

We next investigate the effect of a magnetic impurity on the spin
polarization (SP) of the tunneling
electrons in DMTJ. While the issue of SP for single barriers
has attracted significant interest recently, both theoretically~\cite{tsymbal1,vedyayev}
and experimentally~\cite{moodera2, tsymbal2},
the SP in DMTJ in the presence or absence of
disorder has not been addressed theoretically to the best of our
knowledge.
 The SP for the parallel configuration is
 $SP = (j_P^\uparrow - j_P^\downarrow)/(j_P^\uparrow +
 j_P^\downarrow)$\cite{moodera2, tsymbal2}. In Fig.~3
  we show the SP for the parallel configuration
as a function of $V_{imp}^\uparrow$ for three values of
$V_{imp}^\downarrow$ for $b= 4.3~\AA$ and $7.24~\AA$,
respectively. For b=4.3~\AA, the SP is independent of
$V_{imp}^\uparrow$ due to the fact that the QWS for the majority
spin are antisymmetric and hence $j_P^{\uparrow}$ is not affected
by the impurity. On the other hand, the SP depends on
$V_{imp}^\downarrow$ because the minority spin QWS are symmetric.
In the absence of impurity, $j_P^\uparrow < j_P^\downarrow$ and
hence the SP is negative. An impurity with $V^\downarrow >0$
decreases $j_P^\downarrow$ and hence increases the SP,
leading to a sign reversal of the SP for large enough
$V_{imp}^\downarrow = 2 eV$. In contrast, $j_P^\downarrow$ increases if $V^\downarrow <0 $ and hence the SP decreases.

\begin{figure}
\includegraphics[width=3in,height=3in]{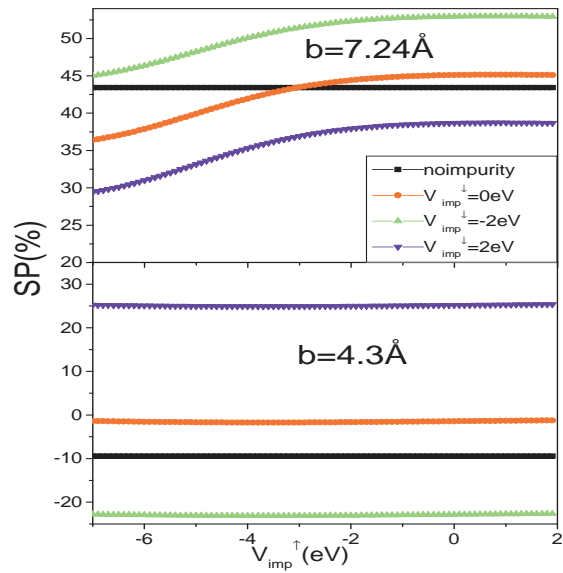}
\caption{\label{fig:epsart} Spin polarization for the parallel
configuration as a function of $V_{imp}^\uparrow$ in the presence
or absence of impurities for different values of
$V_{imp}^\downarrow$ for $b= 4.3~\AA$  and $7.24~\AA$,
respectively.}
\end{figure}

\begin{figure}
\includegraphics[width=3in,height=3in]{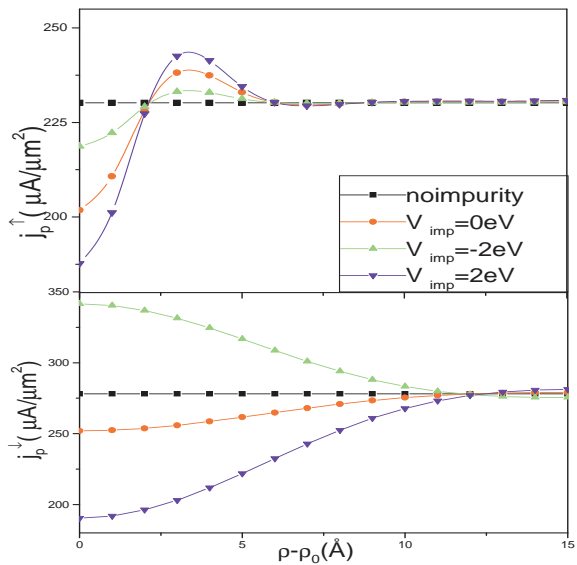}
\caption{\label{fig:epsart} Local current $j^{\sigma}(\rho -
\rho_0)$ in the center of the first barrier as a function of $\rho
- \rho_0$ for $b = 4.3 ~\AA$  and for various values of $V_{imp}$
for the parallel configuration. The nonmagnetic impurity is placed
at the center of the well.}
\end{figure}

For  b= 7.24~\AA, the SP increases with increasing
$V_{imp}^\uparrow$ because the symmetric QWS below E$_F$ are
shifted close to the Fermi energy and $j_P^\uparrow$ increases.
Note that for $V_{imp}^\uparrow > -2eV$ the SP varies weakly with
$V_{imp}^\uparrow$, due to the fact that the QWS lie in an energy range
within k$_B$T around $E_F$. On the other hand, for
$V_{imp}^\downarrow =$ 0 and 2 eV, the QWS states are shifted closer
to $E_F$ ($V^\downarrow > 0$), and hence $j^\downarrow$
increases and the SP decreases. In contrast, for
$V_{imp}^\downarrow = -2 eV$ the QWS states are shifted away
from $E_F$  ($V^\downarrow < 0$), and hence
$j^\downarrow$ decreases and the SP increases.

 In order to understand the
effect of a non-magnetic impurity on the distribution of the local current in
its vicinity  we present in Fig. 4 the local current,
$j^{\sigma}(\rho - \rho_0)$, at the center of the first barrier as
a function of $\rho - \rho_0$ for $b = 4.3~\AA$ and for various
values of $V_{imp}$ for the parallel configuration. The
non-magnetic impurity is placed at the center of the well. It is
interesting to note that while the average current for the
majority spin channel is not affected by the impurity due to the
antisymmetric nature of the QWS (Fig. 2), the local current is dramatically
changed compared to its corresponding value for the pure DMTJ. The
effect is larger as the impurity potential increases. On the other
hand, for the minority spin the local current in the
vicinity of the impurity decreases for $V_{imp} = 0, +2 eV$ and
increases for $V_{imp} = -2 eV$, consistent with the behavior of
the average current behavior (Fig. 1). For distances larger than 10~\AA~the electrons do not feel the impurity.

In conclusion we have presented exact model calculations for the
effect of impurities in the well on the spin-dependent resonant
tunneling for DMTJ. To the best of our knowledge these are the
first calculations to address the effect of impurity in DMTJ. The
calculations reveal that the spin-dependent impurity scattering
potential can tune selectively the majority and minority spin
channels, giving rise to a wide variety of interesting and unusual
transport phenomena. We find that the impurities can lead to a
dramatic enhancement or suppression or sign reversal of the TMR
and the spin polarization. The proposed underlying mechanism which
explains consistently the overall behavior, is the shift of the
original quantum well states depending on (i) the symmetric or
asymmetric nature of these QWS and (ii) the impurity scattering
potential. Our results for the local current indicate that even
though the effect of the impurity on the average spin current is
small for antisymmetric QWS, the local spin current exhibits
strong variation.

We acknowledge useful discussions with J. Moodera. The research at
California State University, Northridge was supported from NSF
under Grant No DMR-00116566 and NASA under Grant No NCC5-513. A.
Vedyayev is grateful to the Russian Foundation of Basic Research
for financial support.

\end{document}